\documentclass[a4paper,fleqn]{cas-dc}

\usepackage[binary-units=true]{siunitx}[=v2]
\usepackage{xurl}
\usepackage{amssymb}
\usepackage[modulo]{lineno} 
\usepackage[ruled, lined, longend,algoruled,boxed]{algorithm2e}
\usepackage[section]{placeins}   

\usepackage{amsthm}

\newdefinition{rmk}{Remark}
\usepackage[numbers]{natbib}
\newcommand{\vect}[1]{\ensuremath{\boldsymbol{\mathrm{#1}}}}

\begin{document}
\let\WriteBookmarks\relax
\def\floatpagepagefraction{1}
\def\textpagefraction{.001}

\shorttitle{RL-Based RTO and Economic NMPC: Experimental Validation}    

\shortauthors{Adhau et al.}  

\title [mode = title]{Two-Timescale Reinforcement Learning for Real-Time Optimization and Economic NMPC: Experimental Validation}

\author[1]{Saket Adhau}[orcid=0000-0003-3776-5675]

\cormark[1]

\ead{saket.adhau@sintef.no}

\credit{Conceptualization, Methodology, Software, Formal analysis, Data Curation, Writing -- Original Draft}

\affiliation[1]{organization={Department of Sustainable Energy Technology, SINTEF Industry},
	addressline={S P Andersens vei $3$}, 
	city={Trondheim},
	postcode={$7031$}, 
	country={Norway}}
\author[4]{Jos\'e Matias}
\ead{jose.matias@kuleuven.be}
\credit{Conceptualization, Methodology, Software, Formal analysis, Data Curation, Writing -- Original Draft}
\affiliation[4]{organization={Department of Chemical Engineering, KU Leuven},
	addressline={Jan Pieter de Nayerlaan $5$},
	city={Sint-Katelijne-Waver},
	postcode={$2860$},
	country={Belgium}}
\author[2]{S\'ebastien Gros}
\ead{sebastien.gros@ntnu.no}
\credit{Conceptualization, Methodology, Formal analysis, Software, Writing -- Review \& Editing, Supervision}
\affiliation[2]{organization={Department of Engineering Cybernetics, Norwegian University of Science and Technology},
	addressline={O. S. Bragstads plass $2$D, Elektroblokk~D},
	city={Trondheim},
	postcode={$7034$}, 
	country={Norway}}

\author[3]{Sigurd Skogestad}
\ead{skoge@ntnu.no}
\credit{Conceptualization, Writing -- Review \& Editing, Supervision, Project administration, Funding acquisition}
\affiliation[3]{organization={Department of Chemical Engineering, Norwegian University of Science and Technology},
	addressline={Sem S\ae landsvei $4$, Kjemiblokk~$5$}, 
	city={Trondheim},
	postcode={$7491$}, 
	country={Norway}}
\begin{abstract}
	We propose a reinforcement learning (RL) framework that tunes both Real-Time Optimization (RTO) and Economic Nonlinear Model Predictive Control (ENMPC) to address plant--model mismatch in process systems. Drawing on modifier-adaptation concepts, the method parameterizes the dynamic model, stage costs, constraints, and RTO modifiers, and uses Q-learning to adjust these parameters at two timescales: a fast update for the ENMPC layer and a slow update for the RTO layer. The framework is experimentally validated on a laboratory rig emulating a three-well subsea oil-production network. Using plant measurement data, the proposed RTO--RLMPC scheme achieves $8.6\%$ higher economic profit than nominal ENMPC, preserves input feasibility of the deployed ENMPC policy and empirically satisfies the path constraints under disturbances and measurement noise, and drives the learned model parameters toward reference plant values, independently identified from experimental data, along the directions that most affect the economic objective. This work provides one of the first experimental demonstrations of RL-tuned ENMPC integrated with RTO.
\end{abstract}

\maketitle
\section{Introduction} \label{sec_introduction}
Process plants must continuously adapt their operating strategies to maximize profit 
or reduce costs under fluctuating market conditions and changing process environments. 
A conventional solution is to decompose decision-making into hierarchical layers 
operating on different time scales 
\citep{skogestad2004control, darby2011rto}. At the top, real-time optimization (RTO) 
uses a nonlinear steady-state model to translate economic objectives into optimal 
setpoints. Below it, model predictive control (MPC) operates on a faster time scale 
with a dynamic model to track these setpoints and reject disturbances, while regulatory 
controllers handle stabilization at the lowest level. Both optimization layers rely on 
process models, and their performance degrades when these models are inaccurate.

When economic objectives and process dynamics are tightly coupled, solving the 
steady-state economic optimization and the dynamic control problem separately, as in 
the conventional hierarchy, can become limiting. Economic 
nonlinear model predictive control (ENMPC) addresses this by integrating economic optimization and 
dynamic control into a single formulation 
\citep{amrit2011economic, rawlings2012fundamentals}. However, the effectiveness of 
both RTO and ENMPC strongly depends on model accuracy. Plant--model mismatch can drive 
the system to suboptimal or infeasible operating points, eroding the expected economic 
benefits \citep{bonvin2013role}. Online parameter estimation can address parametric 
errors, but it is insufficient when the model structure itself is inadequate. 
Discarding the model altogether in favor of a fully data-driven controller avoids 
this issue, but typically requires large amounts of plant data and forfeits the 
structural guarantees, such as constraint satisfaction and interpretability, afforded 
by a model-based formulation. This motivates correcting the model-based optimization 
problem directly from plant measurements while retaining its underlying structure.

One established approach is modifier adaptation (MA) 
\citep{marchetti2009modifier}, which introduces correction terms in the objective and 
constraint functions of the RTO problem: a zeroth-order modifier biases the computed 
constraint to match the measured plant constraint, and a first-order modifier biases 
the computed gradient so that the measured plant gradient is driven to zero. By 
iteratively updating these modifiers from process data, MA can steer the optimization 
toward the true plant optimum, even under significant model inaccuracies 
\citep{marchetti2016modifier}. As the quality of the nominal model deteriorates, MA 
relies increasingly on plant gradient estimates, which may be costly or impractical 
to obtain in practice. An alternative line of work instead uses the 
measured constraints directly, 
avoiding the zeroth-order correction altogether, combined with a model-based estimate 
of the unconstrained gradient rather than the cost 
\citep{krishnamoorthy2021distributed,dirza2022experimental}; 
this reflects the view that satisfying constraints correctly is often more important 
than an accurate gradient.

Reinforcement learning (RL) offers an alternative route: it optimizes closed-loop 
performance directly from plant data, without requiring explicit gradient estimates. 
Recent works \citep{gros2019data} have shown that ENMPC 
schemes can be embedded within an RL framework as structured function approximators, 
where ENMPC serves as a parameterized policy and value function. This retains the 
constraint-aware structure of MPC while enabling data-driven 
performance improvement.

In this work, we build upon these ideas and propose a framework that combines RTO, 
ENMPC, and RL. Inspired by modifier adaptation, we introduce parameterized terms in 
both the RTO and ENMPC layers and use Q-learning to tune them at two timescales: a 
slow update for the RTO parameters and a fast update for the ENMPC parameters. 
Because the Q-learning update is driven by the closed-loop temporal-difference error 
rather than an explicit plant gradient, it eliminates the need for plant gradient 
estimation; and because the slow RTO update targets steady-state economics while the 
fast ENMPC update targets transient dynamics, the two-timescale scheme improves both 
steady-state and transient performance. A natural question is whether this added 
complexity is necessary compared to a single-layer RLMPC. We show experimentally (Section~\ref{sec_results}) that 
the RTO layer improves steady-state profit, while the RL-tuned ENMPC provides 
responsive transient control, and together they outperform simpler alternatives.

The main contributions of this work are:
\begin{enumerate}
	\item A two-timescale RL framework that simultaneously tunes an RTO layer and an 
	ENMPC layer using modifier-adaptation concepts, without requiring plant gradient 
	estimates.
	\item A parameterization of the dynamic model, stage cost, constraints, and RTO 
	modifiers that allows RL to correct for plant--model mismatch at both the 
	steady-state and transient levels.
	\item Experimental validation on a laboratory rig emulating a three-well subsea 
	oil-production network, where the proposed RTO--RLMPC scheme achieves $8.6\%$ higher 
	profit than nominal ENMPC while maintaining input feasibility of the deployed 
	ENMPC policy and empirical path-constraint satisfaction under disturbances and measurement noise.
\end{enumerate}

\section{Background} \label{sec_background}
This section reviews methods for handling plant--model mismatch in 
real-time optimization, the economic NMPC formulation, and the RL-based 
MPC framework of \citet{gros2019data} that forms the basis of our 
approach.

\subsection{RTO and Modifier Adaptation} 
\label{subsec_rto_ma}

A typical RTO problem seeks the economically optimal steady-state 
operating point:
\begin{subequations}
	\begin{align}
		\overline{\vect{x}}, \overline{\vect{u}} 
		~&=~ \arg \min_{\vect{x},\vect{u}} l_p(\vect{x},\vect{u}) \\
		\text{s.t.} \quad 
		\vect{f}(\vect{x},\vect{u}) &= \vect{x} \\
		\vect{h}(\vect{x},\vect{u}) &\leq 0,
	\end{align}
\end{subequations}
where $l_p$ is the economic cost, $\vect{f}$ the steady-state process 
model, and $\vect{h}$ the constraints. In practice, the model 
$\vect{f}$ is derived from first principles and inevitably contains 
errors \citep{darby2011rto}.

The conventional remedy is model parameter adaptation (MPA): the model parameters 
$\vect{\theta}$ are updated so that predictions match plant 
measurements, and the RTO is re-solved with the updated model. If only 
parametric errors are present, MPA may converge to the plant optimum 
\citep{forbes1994model}. However, structural mismatch is almost always 
present in practice, and simply updating parameters of a simplified model 
is rarely sufficient to guarantee optimality 
\citep{quelhas2013common}. To ensure feasibility, 
practitioners introduce safety margins, further reducing the achievable 
economic benefit \citep{quelhas2013common}.

Modifier adaptation (MA) was proposed to overcome this limitation 
\citep{marchetti2009modifier}. Rather than relying solely on parameter 
updates, MA augments the optimization problem with correction terms that 
enforce the KKT conditions of the true plant problem. Assume the plant 
steady-state map is $\vect{y}_p(\vect{u},\vect{d}_p)$, where 
$\vect{u}\in\mathbb{R}^{n_u}$ are the inputs, 
$\vect{d}_p\in\mathbb{R}^{n_d}$ the disturbances, and the available 
model is $\vect{y}(\vect{u},\boldsymbol{\theta})$. MA solves:
\begin{subequations}\label{eq_ma}
	\begin{align}
		\underset{\vect{u}\in\mathbb{U}}{\min} & \quad 
		\phi(\vect{u},\vect{y}(\vect{u},\vect{\theta}_{\mathrm{nom}})) + 
		\boldsymbol{\lambda}_{\phi,k}^T(\vect{u}-\vect{u}_k) \\
		\text{s.t.} & \quad 
		\vect{g}(\vect{u},\vect{y}(\vect{u},\vect{\theta}_{\mathrm{nom}})) 
		+ \boldsymbol{\epsilon}_{g,k} 
		+ \boldsymbol{\lambda}_{g,k}^T(\vect{u}-\vect{u}_k) \leq \vect{0},
	\end{align}
\end{subequations}
where $\phi$ is the economic cost, $\vect{g}$ the constraints, and 
$\mathbb{U}$ the input set. The zeroth-order modifiers 
$\boldsymbol{\epsilon}_{g,k}$ capture the constraint mismatch between 
plant and model:
\begin{equation}
	\boldsymbol{\epsilon}_{g,k} = 
	\vect{g}(\vect{u}_k,\vect{y}_p(\vect{u}_k,\vect{d}_{p,k})) - 
	\vect{g}(\vect{u}_k,\vect{y}(\vect{u}_k,\vect{\theta}_{\mathrm{nom}})) ,
\end{equation}
and the first-order modifiers $\boldsymbol{\lambda}_{\phi,k}$, 
$\boldsymbol{\lambda}_{g,k}$ correct for gradient mismatches:
\begin{subequations}
	\begin{align}
		\boldsymbol{\lambda}_{\phi,k}^T &= 
		\frac{\partial \phi(\vect{u},\vect{y}_p)}{\partial \vect{u}}
		\bigg|_{\vect{u}_k,\vect{d}_{p,k}}
		- \frac{\partial \phi(\vect{u},\vect{y})}{\partial \vect{u}}
		\bigg|_{\vect{u}_k,\vect{\theta}_{\mathrm{nom}}}, \\
		\boldsymbol{\lambda}_{g,k}^T &= 
		\frac{\partial \vect{g}(\vect{u},\vect{y}_p)}{\partial \vect{u}}
		\bigg|_{\vect{u}_k,\vect{d}_{p,k}}
		- \frac{\partial \vect{g}(\vect{u},\vect{y})}{\partial \vect{u}}
		\bigg|_{\vect{u}_k,\vect{\theta}_{\mathrm{nom}}} .
	\end{align}
\end{subequations}
Under mild assumptions, MA converges to the plant optimum by iteratively 
solving \eqref{eq_ma} \citep{marchetti2016modifier}. Extensions combining 
MA with parameter updates have been proposed to accelerate convergence 
\citep{matias2021online}. However, computing the 
first-order modifiers requires plant gradient estimates, which demand 
costly experimental perturbations and hinder practical deployment.
An alternative is to embed the economic objective directly into the 
control layer, avoiding the need for explicit gradient estimation.

\subsection{Economic Nonlinear MPC}
\label{subsec_enmpc}

Economic NMPC (ENMPC) removes the separation between the RTO and control 
layers by optimizing an economic objective directly over a dynamic 
prediction horizon \citep{amrit2011economic}. The 
ENMPC problem reads:
\begin{subequations}\label{eq_enmpc}
	\begin{align}
		\min_{\{\vect{x}_k\}_{k=0}^{N},\,\{\vect{u}_k\}_{k=0}^{N-1}} \quad
		& \sum_{k=0}^{N-1} \Phi(\vect{u}_k,\vect{y}_k) + V_f(\vect{x}_N) \\
		\text{s.t.}\quad
		 \vect{x}_{k+1} &= F(\vect{x}_k,\vect{u}_k,\vect{\theta}_{\mathrm{nom}}), 
		\quad k=0,\dots,N-1, \\
		\vect{y}_k &= H(\vect{x}_k,\vect{u}_k), 
		\quad k=0,\dots,N-1, \\
		 \vect{x}_0 &= \vect{x}(0), \\
		 G(\vect{u}_k,\vect{y}_k) &\le  0, \quad k=0,\dots,N-1, \\
		 G_f(\vect{x}_N) &\le  0, 
	\end{align}
\end{subequations}
where $\vect{x}$ are the states, $\vect{y}$ the measured outputs, 
$F$ and $H$ the nonlinear model, $\Phi$ the economic stage cost, 
$V_f$ the terminal cost, and $G$, $G_f$ the stage and terminal 
constraints. Terminal conditions are imposed to ensure closed-loop 
stability \citep{rawlings2012fundamentals}. Since ENMPC relies on a fixed 
nominal model, its performance degrades under plant--model mismatch, and 
feedback alone is generally insufficient to recover optimality 
\citep{bonvin2013role}. To address this, one can leverage data-driven 
methods to adapt the controller online.

\subsection{RL-Based MPC} \label{sec_rl_mpc}

Reinforcement learning (RL) provides a framework for optimizing 
closed-loop control policies directly from plant data 
\citep{sutton2018reinforcement,bertsekas2019reinforcement}. Given a policy 
$\vect{\pi}$ mapping states $\vect{s}_k$ to actions 
$\vect{a}_k = \vect{\pi}(\vect{s}_k)$, the goal is to find the policy 
minimizing the expected discounted cost:
\begin{align}
	\vect{\pi}^\star = 
	\arg\min_{\vect{\pi}}\, 
	\mathbb{E}\!\left[
	\sum_{k=0}^\infty \gamma^k 
	L(\vect{s}_k,\vect{a}_k)
	\,\Bigg|\,
	\vect{a}_k = \vect{\pi}(\vect{s}_k)
	\right],
\end{align}
where $L(\vect{s}_k,\vect{a}_k)$ is the stage cost and 
$\gamma \in (0,1]$ the discount factor. The optimal policy satisfies 
the Bellman equations:
\begin{subequations}
	\begin{align}
		Q^\star(\vect{s},\vect{a}) &= 
		L(\vect{s},\vect{a}) + 
		\gamma \, \mathbb{E}[V^\star(\vect{s}_+) \mid \vect{s},\vect{a}], \\
		V^\star(\vect{s}) &= 
		\min_{\vect{a}} Q^\star(\vect{s},\vect{a}),
	\end{align}
\end{subequations}
where $Q^\star$ is the optimal action--value function, $V^\star$ the 
optimal value function, and $\vect{s}_+$ the successor state.

A key challenge is the choice of function approximators for $Q^\star$ 
and $\vect{\pi}^\star$. Generic approximators such as deep neural 
networks are flexible but lack guarantees on constraint satisfaction and 
closed-loop stability. To address this, \citet{gros2019data} proposed 
using a parameterized ENMPC scheme as the function approximator, 
embedding process knowledge and MPC theory directly into the RL 
framework. In this construction, the terminal cost $V_{f,\theta}$ 
and a stage-cost lower bound by a $\mathcal{K}_\infty$ function play 
a dual role: they bridge classical ENMPC dissipativity-based stability 
arguments with the matching condition that allows the parameterized 
scheme to recover the optimal action--value function under 
plant--model mismatch \citep{gros2019data}. The parameterized ENMPC 
problem is:
\begin{subequations}\label{eq_enmpc_rl}
	\begin{align}
		\min_{\{\vect{x}_k\},\,\{\vect{u}_k\}} \quad 
		& \gamma^N V_{f,\theta}(\vect{x}_N) + 
		\sum_{k=0}^{N-1} \gamma^k\,\ell_\theta(\vect{x}_k,\vect{u}_k) \\
		\text{s.t.} \quad 
		& \vect{x}_{k+1} = f_\theta(\vect{x}_k,\vect{u}_k), 
		\quad k=0,\dots,N-1, \\
		& h_\theta(\vect{x}_k,\vect{u}_k) \leq 0, 
		\quad k=0,\dots,N-1 ,
	\end{align}
\end{subequations}
where $\ell_\theta$ and $V_{f,\theta}$ are parameterized stage and 
terminal costs, $f_\theta$ the parameterized dynamics, and $h_\theta$ 
parameterized constraints. We use lowercase notation ($f_\theta$, $\ell_\theta$, $h_\theta$) to distinguish the parameterized components from their nominal counterparts ($F$, $\Phi$, $G$) in \eqref{eq_enmpc}. Compared with \eqref{eq_enmpc}, the explicit terminal inequality 
$G_f(\vect{x}_N)\le 0$ is absorbed into the parameterized terminal cost 
$V_{f,\theta}(\vect{x}_N)$, following the terminal-cost ENMPC framework 
of \citet{amrit2011economic}. The parameter vector $\vect{\theta}$ collects all 
tunable parameters appearing in $f_\theta$, $\ell_\theta$, $V_{f,\theta}$, and 
$h_\theta$, i.e., the dynamics, stage cost, terminal cost, and constraints, and 
thereby defines the policy and value function representation. We identify the RL state with 
the plant state, $\vect{s} \equiv \vect{x}$, and the RL action with the 
control input, $\vect{a} \equiv \vect{u}$.

The policy is given by the first element of the optimal input sequence:
\begin{align}
	\vect{\pi}_\theta(\vect{s}) = \vect{u}_0^\star ,
\end{align}
and the action--value function by:
\begin{align}\label{eq_av_parametrized}
	Q_\theta(\vect{s},\vect{a}) = 
	\min_{\{\vect{x}_k\},\,\{\vect{u}_k\}}\;\,
	&\gamma^N V_{f,\theta}(\vect{x}_N) \notag \\
	&+ \sum_{k=0}^{N-1} \gamma^k\,\ell_\theta(\vect{x}_k,\vect{u}_k),
\end{align}
subject to \eqref{eq_enmpc_rl} with $\vect{u}_0 = \vect{a}$. The value 
function is:
\begin{align}\label{eq_vf_parameterization}
	V_\theta(\vect{s}) 
	&= \min_{\vect{a}} Q_\theta(\vect{s},\vect{a}) .
\end{align}
These parameterizations are consistent with the Bellman equations by 
construction. The role of RL is to tune $\theta$ from closed-loop data 
so that the ENMPC scheme approximates $\vect{\pi}^\star$, even under 
plant--model mismatch.

\section{Proposed RTO--RLMPC Framework} 
\label{sec_proposed_method}

The parameterized ENMPC formulation in \citet{gros2019data} can generate 
an optimal policy even when the model is inaccurate, provided the 
parameterization is sufficiently rich. However, that formulation focuses 
primarily on transient performance and does not explicitly account for 
steady-state economics. As a result, the closed-loop trajectories may 
converge to a steady state that differs from the true plant optimum 
\citep{vaccari2017modifier}.

To address this, we propose to parameterize both the RTO and ENMPC 
layers. The RTO layer is augmented with modifier terms so that the 
computed equilibrium corresponds to the true plant optimum, and the 
ENMPC layer is parameterized in its model, stage cost, terminal cost, 
and constraints. RL is then used to tune the parameters across both 
layers at two timescales: ENMPC parameters are updated at every control 
step (fast), while RTO modifier parameters are updated once per episode 
(slow). This eliminates the need for plant gradient estimation while 
ensuring consistent adaptation across layers.

Alternative designs are less effective: a parameterized RTO with a 
conventional tracking MPC limits the ENMPC layer to setpoint tracking 
and forgoes transient economic optimization; a parameterized ENMPC 
with a standard RTO leaves the RTO layer unable to adapt to mismatch; 
and tuning RTO and ENMPC independently with RL doubles the learning 
complexity and fails to ensure consistency between layers.

\noindent\textbf{Sign conventions.} The RTO layer in \eqref{eq_rto_param} is stated as a profit \emph{maximization}, while the RL formulation and the ENMPC stage cost are stated as \emph{minimizations}. We treat the RL stage cost $L$ as the negative of the instantaneous economic profit so that the two conventions are consistent throughout.

\subsection{Parameterized RTO with Modifier Terms}
\label{subsec_param_rto}

We build on the modifier-adaptation framework to construct a 
parameterized RTO scheme. For detailed discussions of modifier-adaptation 
the reader is referred to \citep{marchetti2009modifier,marchetti2016modifier}. 

Consider the nonlinear process model
\begin{subequations}\label{eq_rto_model}
	\begin{align}
		\vect{x}_{k+1} &= f_{\theta}(\vect{x}_k,\vect{u}_k), \\
		\vect{y}_k &= H(\vect{x}_k,\vect{u}_k),
	\end{align}
\end{subequations}
where $\vect{u}_k \in \mathbb{R}^{n_u}$ are the decision variables, 
$\vect{y}_k \in \mathbb{R}^{n_y}$ the measured outputs, and 
$\vect{x}_k \in \mathbb{R}^{n_x}$ the states. The dynamics $f_\theta$ 
are parameterized by $\vect{\theta} \in \mathbb{R}^{n_m}$, so that 
$\vect{y}_k$ inherits a $\vect{\theta}$-dependence indirectly through 
$\vect{x}_k(\vect{\theta})$; the output map $H$ itself is known and 
parameter-free.

At each iteration $k$, the RTO problem is formulated as
\begin{subequations}\label{eq_rto_param}
	\begin{align}
		\vect{u}^\star_{k+1}, \vect{y}^\star_{k+1} 
		=~ &\arg \max_{\vect{u}} \; 
		\bar{\Phi}(\vect{y},\vect{u}) 
		+ (\vect{\lambda}_{\theta}^{\phi})_k^{\top}(\vect{u}-\vect{u}_k) \\
		\text{s.t.} \quad 
		& \vect{y} = f_{\mathrm{ss},\theta}(\vect{u}), \\
		& \underline{\vect{u}} \leq \vect{u} \leq \overline{\vect{u}}, \\
		& \vect{g}_{\theta}(\vect{y},\vect{u}) + (\vect{\epsilon}^{g}_{\theta})_k 
		+ (\vect{\lambda}_{\theta}^{g})_k^{\top}(\vect{u}-\vect{u}_k) \leq 0 ,
	\end{align}
\end{subequations}
where $\bar{\Phi}$ is the economic profit, $\vect{g}_{\theta}$ are nonlinear 
constraints, and $\underline{\vect{u}}, \overline{\vect{u}}$ denote input 
bounds. The additional terms $(\vect{\lambda}_{\theta}^{\phi})^{\top}$, 
$(\vect{\lambda}_{\theta}^{g})^{\top}$, and $\vect{\epsilon}^g_\theta$ act as modifier 
terms to correct for plant--model mismatch. If the true cost or 
constraints are exactly known functions of $\vect{u}$, the corresponding 
corrections vanish.

These modifier terms closely parallel those in conventional modifier 
adaptation \citep{forbes1994model}, 
where they are computed from plant gradients. For example, at $\vect{u}_k$ 
the plant gradients 
$\tfrac{\partial \bar{\Phi}_p}{\partial \vect{u}}$ and 
$\tfrac{\partial g_p}{\partial \vect{u}}$ are typically required. In the 
proposed framework, however, reinforcement learning eliminates the need 
for explicit gradient estimation by updating modifier terms directly from 
observed input--output trajectories.

\begin{figure}[]
	\centering
	\includegraphics[width=\columnwidth]{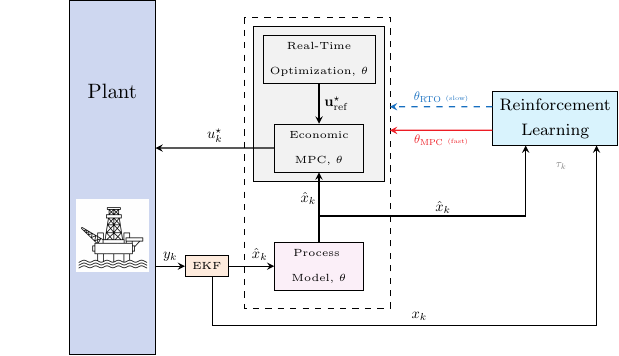}
	\caption{Schematic of the proposed framework: reinforcement learning (RL) updates 
		the modifier terms of the real-time optimization (RTO) and the economic 
		nonlinear MPC (ENMPC) layers, ensuring consistent adaptation and improved 
		closed-loop economic performance.}
	\label{fig_mainidea}
\end{figure}

The computed optimal input $\vect{u}^\star_{k+1}$ from 
\eqref{eq_rto_param} is then passed to the ENMPC layer as a reference 
embedded in the stage-cost tracking term of $\ell_\theta$, penalizing 
deviation of $\vect{u}_k$ from $\vect{u}^\star_{k+1}$; asymptotic 
stability then follows from standard terminal-cost ENMPC arguments 
\citep{amrit2011economic}. 
We partition the full parameter vector as 
$\vect{\theta} = (\vect{\theta}_V,\vect{\theta}_m,\vect{\theta}_c,\vect{\theta}_{\mathrm{RTO}})$: 
$\vect{\theta}_V$ is the terminal-cost parameter, $\vect{\theta}_m$ the 
model parameters, $\vect{\theta}_c$ the stage-cost and constraint weights, 
and $\vect{\theta}_{\mathrm{RTO}}$ the RTO modifier parameters; this 
partition is defined in detail in Section~\ref{subsec_qlearning}. 
Fig.~\ref{fig_mainidea} illustrates the overall framework; the complete 
procedure (Algorithm~\ref{alg_main}) is given at the end of 
Section~\ref{subsec_sensitivities}, once the sensitivities it relies on 
have been introduced.

To illustrate the role of the model in RLMPC \citep{gros2019data}, 
consider two extreme cases: with no or minimal model, RL may still 
converge to the plant optimum given a sufficiently rich parameterization 
and enough data, behaving similarly to extremum-seeking control, which 
optimizes directly from measured plant response without relying on a 
process model \citep{ariyur2003realtime}; with a perfect model, the 
parameterized correction terms vanish and plant optimality is achieved 
directly. In practice, the available model captures the main dynamics 
but is insufficient for plant optimality, and RL updates the 
parameterized terms so that the model becomes locally accurate for 
optimization at each iteration, progressing in small steps toward the 
plant optimum.

\subsection{$Q$-Learning for ENMPC} 
\label{subsec_qlearning}

Classical $Q$-learning provides a mechanism to adjust 
the parameters $\vect{\theta}$ of the parameterized ENMPC scheme 
\eqref{eq_vf_parameterization}. In its basic form, $Q$-learning minimizes 
the temporal-difference (TD) error:
\begin{subequations}\label{eq_q_learning}
	\begin{align}
		\tau_k &= L(\vect{s}_k,\vect{a}_k) 
		+ \gamma V_\theta(\vect{s}_{k+1}) 
		- Q_\theta(\vect{s}_k,\vect{a}_k), 
		\label{eq_q_learning_td_update}
	\end{align}
	where $Q_\theta(\vect{s}_k,\vect{a}_k)$ and $V_\theta(\vect{s}_{k+1})$ are recovered 
	by solving the parameterized ENMPC problem \eqref{eq_av_parametrized} at 
	$(\vect{s}_k,\vect{a}_k)$ and \eqref{eq_vf_parameterization} at $\vect{s}_{k+1}$, 
	respectively, and the stage cost $L$ is evaluated using the \emph{plant} 
	measurements:
	\begin{align}\label{eq_q_learning_baseline_cost}
		L(\vect{s}_k,\vect{a}_k) = 
		\ell_{\theta}(\vect{x}_k,\vect{u}_k) + 
		\vect{w}^\top \max\!\left(0,\,h_\theta(\vect{x}_k,\vect{u}_k)\right),
	\end{align}
	with $\vect{x}_k$ obtained from plant measurements (via the state 
	estimator) and $\vect{u}_k$ chosen according to the ENMPC policy 
	$\vect{\pi}_{\theta}(\vect{s}_k)$. The vector 
	$\vect{w} \in \mathbb{R}^{n_h}_{\ge 0}$ collects componentwise 
	penalty weights for the $n_h$ inequalities in $h_\theta$; its 
	entries are fixed (not learned) and sized so that persistent 
	constraint violations are steadily driven out of the learned 
	parameters, without dominating the economic objective on typical 
	trajectories \citep{gros2019data}. The numerical values used in the case study 
	are given in Section~\ref{subsec_problem_formulation}. The 
	parameters are updated as
	\begin{align}\label{eq_q_learning_theta_update}
		\vect{\theta} \leftarrow 
		\vect{\theta} + \alpha \tau_k \frac{d Q_\theta}{d \theta}\bigg|_{\vect{x}_k,\vect{u}_k},
	\end{align}
	where $\alpha > 0$ is the learning rate.
\end{subequations}

The evaluation of $\tfrac{d Q_\theta}{d \theta}$ requires the sensitivities of 
the action--value function in \eqref{eq_av_parametrized}, as discussed in 
\citet{gros2019data}. Recall the partition 
$\vect{\theta} = (\vect{\theta}_V,\vect{\theta}_m,\vect{\theta}_c,\vect{\theta}_{\mathrm{RTO}})$ 
introduced in Section~\ref{subsec_param_rto}: $\vect{\theta}_V$ collects the terminal-cost parameters and 
$\vect{\theta}_c$ the stage-cost/constraint weights (both enter only 
the ENMPC \eqref{eq_enmpc_rl}); $\vect{\theta}_m$ are model parameters 
shared by the ENMPC and the RTO \eqref{eq_rto_param}; and 
$\vect{\theta}_{\mathrm{RTO}}$ are the modifier parameters 
$(\vect{\lambda}^\phi_\theta,\vect{\lambda}^g_\theta,\vect{\epsilon}^g_\theta)$ 
appearing only in the RTO. Denote by $\vect{\theta}_{\mathrm{sub}} = 
(\vect{\theta}_m,\vect{\theta}_{\mathrm{RTO}})$ the subset of parameters 
that actually enter the RTO problem. In the proposed RTO--RLMPC 
framework, $\vect{\theta}$ 
affects $Q_\theta$ both directly through the ENMPC scheme and indirectly 
via the RTO problem \eqref{eq_rto_param}, since the optimal steady-state 
input $\vect{u}^\star$ enters the ENMPC through the reference-tracking 
term of the stage cost $\ell_\theta$. The resulting total derivative is therefore
\begin{align}\label{eq_total_derivative}
	\frac{d Q_\theta}{d \theta} = 
	\frac{\partial Q_\theta}{\partial \theta} 
	+ \frac{\partial Q_\theta}{\partial \vect{u}^\star} 
	\frac{\partial \vect{u}^\star}{\partial \vect{\theta}_{\mathrm{sub}}},
\end{align}
where $\tfrac{\partial \vect{u}^\star}{\partial \vect{\theta}_{\mathrm{sub}}}$ 
is the block sensitivity extracted from \eqref{eq_rto_sensitivities}, 
and the corresponding blocks for $(\vect{\theta}_V,\vect{\theta}_c)$ are 
identically zero.

In practice, we adopt a two-timescale update strategy: the ENMPC 
parameters $\vect{\theta}$ are updated at every time step $k$ using the 
instantaneous TD error and gradient, while the RTO modifier parameters 
$\vect{\theta}_{\mathrm{RTO}}$ are updated once per episode using the 
episode-averaged TD error and sensitivities 
$\tfrac{\partial \vect{u}^\star}{\partial \vect{\theta}_{\mathrm{RTO}}}$, 
since the RTO layer targets steady-state economics and should evolve 
slowly to avoid oscillations, whereas the ENMPC layer handles transient 
dynamics and benefits from faster adaptation. Rather than triggering a 
new RTO iteration via an explicit steady-state test on process data, as 
in conventional RTO \citep{darby2011rto}, the episode length $T$ is 
fixed \emph{ad hoc} from process knowledge of the closed-loop settling 
time; this avoids the lengthy waiting procedure of conventional MA, 
where a new iteration is withheld until the plant has settled and plant 
gradients have been estimated, but introduces a trade-off: too short a 
$T$ biases the episode-averaged sensitivities and TD error with 
transient data, while too long a $T$ slows RTO adaptation and wastes 
data. This design is inspired by two-timescale stochastic approximation 
theory \citep{borkar1997stochastic}; the extent to which the classical 
convergence guarantees apply is discussed in 
Remark~\ref{rmk_convergence}.

\begin{rmk}[Convergence of the two-timescale scheme]\label{rmk_convergence}
The classical two-timescale stochastic approximation framework of 
\citet{borkar1997stochastic} requires (i) diminishing step sizes with 
$\sum_k \alpha_k = \infty$, $\sum_k \alpha_k^2 < \infty$, and 
$\alpha_{\mathrm{slow},k}/\alpha_{\mathrm{fast},k} \to 0$; (ii) 
boundedness of the iterates; (iii) Lipschitz continuity of the update 
maps; and (iv) global asymptotic stability of the limiting ODEs. In the 
proposed framework, condition~(i) is not met as stated: constant step 
sizes are used for both layers, so the Robbins--Monro conditions are 
violated, and the update-frequency separation (one ENMPC update per 
control step, one RTO update per episode of length $T$) only 
\emph{heuristically} mimics the required ratio condition through the 
effective per-episode ratio $\alpha_{\mathrm{slow}}/(T\,\alpha_{\mathrm{fast}})$; 
under constant step sizes only weaker convergence-in-distribution 
guarantees are available \citep{borkar1997stochastic}, and formally 
establishing these for the present NLP-based setting is left for future 
work. Condition~(ii) holds because the cost-weight matrices are 
projected onto the positive-semidefinite cone at each episode and the 
model parameters are physically bounded. Condition~(iii) holds 
piecewise, since the NLP sensitivities $\partial Q_\theta/\partial 
\theta$ are smooth between active-set changes under LICQ and SOSC 
\citep{gros2019data}, though discontinuities may occur at transitions. 
Condition~(iv) remains an open theoretical question, though empirically 
the TD error converges to zero within approximately $150$ episodes 
(Fig.~\ref{fig_td_convergence}) and the learned parameters stabilize, 
suggesting well-behaved limiting dynamics for the present case study.
\end{rmk}

In the next subsection, we outline how these sensitivities are computed 
for \eqref{eq_av_parametrized} and \eqref{eq_rto_param}.

\begin{algorithm}[t] 
	\caption{RTO--RLMPC framework with $Q$-learning}
	\label{alg_main}
	\SetKwInOut{Input}{Input}
	\SetKwInOut{Output}{Output}
	
	\Input{initial $\vect{\theta}_0=(\vect{\theta}_{V,0},\vect{\theta}_{m,0},\vect{\theta}_{c,0},\vect{\theta}_{\mathrm{RTO},0})$ 
		(terminal-cost, model, stage-cost/constraint, and RTO-modifier parameters, respectively), 
		initial state $\vect{x}_0$, exploration $\sigma_0,\sigma_{\min},\rho$, 
		learning rates $\alpha,\alpha_m,\alpha_R$, total-input limit $\bar u_{\mathrm{tot}}$ 
		(application-specific; e.g., the total gas availability of Section~\ref{subsec_problem_formulation}), 
		model-parameter bounds $[\underline{\vect\theta}_m,\overline{\vect\theta}_m]$, 
		rejection limit $N_{\max}$}
	\Output{learned parameters $\vect{\theta}^\star$}
	
	\For{episode $k=1,\dots,K$}{
		$\sigma_k \leftarrow \max(\sigma_{\min},\sigma_0\rho^{k-1})$; solve RTO \eqref{eq_rto_param} for target $\vect{u}^\star$ and sensitivity $\partial\vect{u}^\star/\partial\vect{\theta}_{\mathrm{sub}}$ via \eqref{eq_rto_sensitivities}\;
		Initialize policy from $V_\theta(\vect{x}_0)$ \eqref{eq_vf_parameterization}\;
		\For{$i=1,\dots,T$}{
			$\vect{a}_i \leftarrow \mathrm{sat}(\vect{\pi}_\theta(\vect{x}_i)+\vect e,\underline{\vect u},\overline{\vect u})$ with $\vect e\sim\mathcal{N}(0,\sigma_k^2 I)$; 
			enforce the application-specific shared-input $\sum_j a_{i,j}\le \bar u_{\mathrm{tot}}$ 
			(a coupling constraint additional to the box constraints in $h_\theta$; see Section~\ref{subsec_problem_formulation}) 
			by rejection sampling ($N_{\max}$ attempts) then rescaling if needed\;
			Apply $\vect{a}_i$; observe $\vect{x}_{i+1}$, stage cost $L(\vect{x}_i,\vect{a}_i)$ \eqref{eq_q_learning_baseline_cost}\;
			Solve $Q_\theta$ \eqref{eq_av_parametrized} and $V_\theta(\vect{x}_{i+1})$ \eqref{eq_vf_parameterization}; get $\tfrac{dQ_\theta}{d\theta}$ via \eqref{eq_enmpc_sensitivities}--\eqref{eq_total_derivative}\;
			$\tau_i \leftarrow L(\vect{x}_i,\vect{a}_i)+\gamma V_\theta(\vect{x}_{i+1})-Q_\theta(\vect{x}_i,\vect{a}_i)$ \eqref{eq_q_learning_td_update}\;
			$(\vect{\theta}_V,\vect{\theta}_c) \leftarrow (\vect{\theta}_V,\vect{\theta}_c) + \alpha\,\tau_i\,\tfrac{\partial Q_\theta}{\partial(\vect{\theta}_V,\vect{\theta}_c)}$;\;
			$\vect{\theta}_m \leftarrow \vect{\theta}_m + \alpha_m\,\tau_i\,\tfrac{\partial Q_\theta}{\partial \vect{\theta}_m}$\;
			Project the weight matrices within $\vect{\theta}_c$ onto $\succeq 0$ (concretely $R,Q_u$; Section~\ref{subsec_process_modeling}); clip $\vect{\theta}_m$ to $[\underline{\vect\theta}_m,\overline{\vect\theta}_m]$\;
		}
		$\bar\tau_k \leftarrow \tfrac{1}{T}\sum_i \tau_i$;\quad 
		$\vect{\theta}_{\mathrm{RTO}} \leftarrow \vect{\theta}_{\mathrm{RTO}} + \alpha_R\,\bar\tau_k\,\tfrac{1}{T}\sum_i \tfrac{dQ_\theta}{d\vect{\theta}_{\mathrm{RTO}}}$\;
	}
\end{algorithm}
\subsection{Sensitivities for the RTO--RLMPC Update}
\label{subsec_sensitivities}

To compute the sensitivities required in the parameter update, we denote 
the Lagrangian functions associated with the ENMPC problem 
\eqref{eq_av_parametrized} and the RTO problem \eqref{eq_rto_param} as 
$\mathcal{L}_\theta$ and $\bar{\mathcal{L}}_\theta$, respectively:
\begin{subequations}
	\begin{align}
		\mathcal{L}_\theta &= 
		\Phi_\theta + \lambda^{\top} C_\theta + \mu^{\top} G_\theta, \\ 
		\bar{\mathcal{L}}_\theta &= 
		\bar{\Phi}_\theta + \bar{\lambda}^{\top} \bar{C}_\theta 
		+ \bar{\mu}^{\top} \bar{G}_\theta,
	\end{align}
\end{subequations}
where $C_\theta$, $G_\theta$ denote the equality and inequality 
constraints of the ENMPC problem, and $\bar{C}_\theta$, $\bar{G}_\theta$ 
those of the RTO problem. The primal--dual solution vectors are 
$\vect{z}=\{\vect{p},\lambda,\mu\}$ and 
$\bar{\vect{z}}=\{\bar{\vect{p}},\bar{\lambda},\bar{\mu}\}$, 
respectively.

Under standard regularity (KKT holds, LICQ + SOSC, strict complementarity), 
the ENMPC sensitivity satisfies \citep{gros2019data}
\begin{align}\label{eq_enmpc_sensitivities}
	\frac{\partial Q_\theta}{\partial \theta} \;=\; 
	\frac{\partial \mathcal{L}_\theta(\vect{z}^\star)}{\partial \theta},
\end{align}
where $\vect{z}^\star$ is the optimal primal--dual solution of 
\eqref{eq_av_parametrized}.

For the RTO NLP \eqref{eq_rto_param}, define the KKT residual
\begin{align}
	\mathcal{R}_\theta(\bar{\vect{z}}) =
	\begin{bmatrix}
		\nabla_{\bar{\vect{p}}}\,\bar{\mathcal{L}}_\theta \\
		\bar{C}_\theta \\
		\operatorname{Diag}(\bar{\mu})\,\bar{G}_\theta
	\end{bmatrix} = \vect{0},
	\quad \bar{G}_\theta \le 0,\; \bar{\mu} \ge 0.
\end{align}
By the implicit function theorem,
\begin{subequations}\label{eq_rto_sensitivities}
	\begin{align}
		\frac{\partial \bar{\vect{z}}^\star}{\partial \theta} = 
		-\left( \frac{\partial \mathcal{R}_\theta}{\partial \bar{\vect{z}}} 
		\right)^{-1} 
		\frac{\partial \mathcal{R}_\theta}{\partial \theta}
		\Bigg|_{\bar{\vect{z}}=\bar{\vect{z}}^\star},
	\end{align}
\end{subequations}
and since $\vect{u}^\star$ is a component of $\bar{\vect{z}}^\star$, the 
needed block sensitivity 
$\tfrac{\partial \vect{u}^\star}{\partial \vect{\theta}_{\mathrm{sub}}}$ 
with $\vect{\theta}_{\mathrm{sub}} = (\vect{\theta}_m,\vect{\theta}_{\mathrm{RTO}})$ 
is extracted from $\tfrac{\partial \bar{\vect{z}}^\star}{\partial \vect{\theta}_{\mathrm{sub}}}$; 
the corresponding blocks for $(\vect{\theta}_V,\vect{\theta}_c)$ vanish 
because these parameters do not appear in \eqref{eq_rto_param}.

With all required sensitivities now in place, the complete two-timescale 
learning procedure is summarized in Algorithm~\ref{alg_main}.

\section{Case Study: Subsea Gas-Lift Oil-Well Network} \label{sec_case_study}
\begin{figure}[]
	\centering
	\includegraphics[width=\columnwidth]{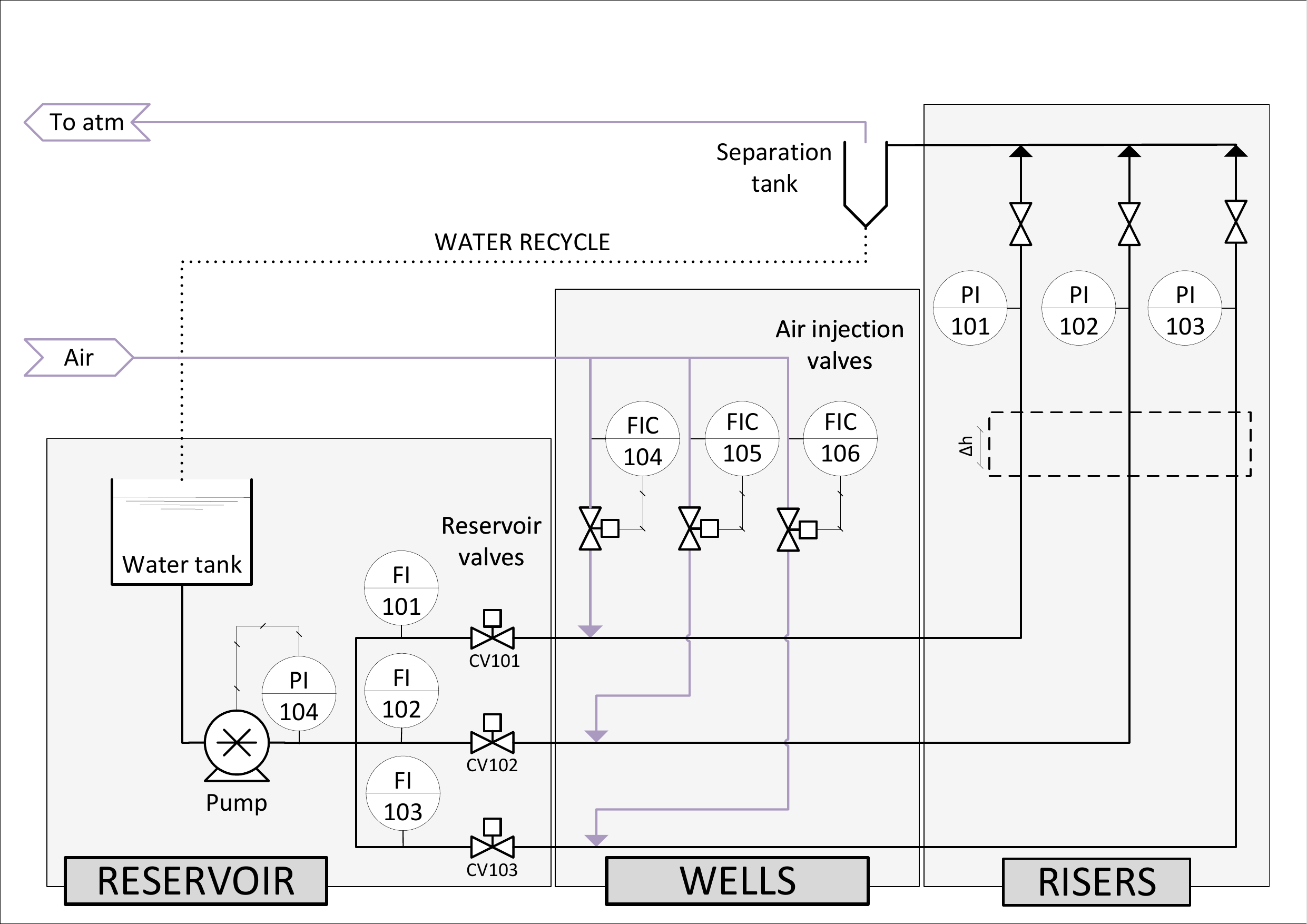}
	\caption{Experimental schematic adapted from \citep{matias2021steady}. Gas 
		injection flow rates (FI$104$, FI$105$, FI$106$) serve as manipulated 
		inputs. Measured outputs include wellhead pressures (PI$101$--PI$103$), 
		pump outlet pressure (PI$104$), and liquid flow rates (FI$101$--FI$103$). 
		Reservoir valve openings (CV$101$--CV$103$) act as disturbances.}
	\label{fig_flowsheet}
\end{figure}

The proposed framework is validated on a laboratory rig emulating a 
three-well subsea oil-production network with gas lift 
\citep{matias2021steady}. In gas-lift systems, compressed gas is injected 
into the well to reduce the hydrostatic pressure and increase liquid 
production \citep{amara2016gas}. However, excessive injection raises 
frictional losses and ultimately reduces output 
\citep{hernandez2016fundamentals}. The optimization problem is therefore 
to allocate a limited supply of lift gas among the wells so as to 
maximize total oil production. A schematic of the experimental rig is 
shown in Fig.~\ref{fig_flowsheet}.

\subsection{Rig Description} \label{subsec_exp_rig}

The rig, previously described in \citep{matias2021steady}, employs air 
and water as substitutes for natural gas and crude oil, a common, 
dynamics-preserving substitution in experimental studies of subsea oil 
wells \citep{jahanshahi2017anti}.

The system is divided into three sections: reservoir, wells, and riser, 
summarized in Table~\ref{tab_rig_specs}.

\begin{table}[]
	\centering
	\caption{Physical specifications of the experimental rig sections 
		\citep{matias2021steady}.}
	\label{tab_rig_specs}
	\begin{tabular}{lp{0.72\columnwidth}}
		\hline
		Section & Specification \\
		\hline
		Reservoir & $200\,\si{\liter}$ stainless-steel tank with a 
		centrifugal pump; outlet pressure (PI$104$) held at 
		$0.3\,\si{\bar}$; valves CV$101$--CV$103$ emulate reservoir 
		disturbances; liquid outflow $2$--$15\,\si{\liter\per\minute}$ 
		(FI$101$--FI$103$). \\
		Wells & Three flexible hoses, $2\,\si{\centi\meter}$ inner 
		diameter, $1.5\,\si{\meter}$ length; air injected 
		$\approx 10\,\si{\centi\meter}$ downstream of the reservoir 
		valves via flow controllers FIC$104$--FIC$106$, range 
		$1$--$5\,\si{s\liter\per\minute}$. \\
		Riser & Three vertical pipes, $2\,\si{\centi\meter}$ inner 
		diameter, $2.2\,\si{\meter}$ length, orthogonal to the wells; 
		wellhead pressure measured at the riser top (PI$101$--PI$103$), 
		followed by manual valves. \\
		\hline
	\end{tabular}
\end{table}

\subsection{State Estimation} \label{subsec_state_estimation}

Since not all states are directly measurable, an Extended Kalman Filter 
(EKF) is employed to estimate the full state vector from available 
measurements. The EKF operates on the nonlinear DAE model described in 
Appendix~\ref{appendix_model_eq} and uses the measured wellhead pressures 
(PI$101$--PI$103$) and liquid flow rates (FI$101$--FI$103$) as outputs. 
The estimated states are then passed to the ENMPC and RTO solvers at each 
sampling instant. The EKF tuning was performed offline, and the noise covariance matrices were calibrated from the 
experimental noise characteristics of the rig instrumentation.

\subsection{Process Modeling and Parameterization} 
\label{subsec_process_modeling}

The first-principles dynamic model of the rig was developed in 
\citep{matias2021steady} and is summarized in 
Appendix~\ref{appendix_model_eq}. The proposed methodology can adapt to 
structural mismatch provided that certain model parameters are 
adjustable. These parameters should be selected to capture the dominant 
process disturbances. In the present rig, the main disturbances originate from the reservoir valves (CV$101$--CV$103$). 
Accordingly, we introduce parameter updates associated with the valve 
behavior. The set of adjustable parameters is
\begin{align}
	\vect{\theta}_m = 
	\begin{bmatrix} 
		\theta_{\mathrm{res},1} & \theta_{\mathrm{res},2} & \theta_{\mathrm{res},3} & 
		\theta_{\mathrm{top},1} & \theta_{\mathrm{top},2} & \theta_{\mathrm{top},3} 
	\end{bmatrix}^{\top},
\end{align}
where $\theta_{\mathrm{res},i}$ are reservoir valve coefficients and 
$\theta_{\mathrm{top},i}$ denote corresponding top-pressure parameters, with 
$i=1,2,3$ indexing the wells.

\noindent\textbf{Concrete ENMPC parameterization.} 
The generic parameterized objects $\ell_\theta,\,V_{f,\theta},\,h_\theta$ 
in \eqref{eq_enmpc_rl} are instantiated as
\begin{subequations}\label{eq_param_instantiation}
\begin{align}
	\ell_\theta(\vect x,\vect u,\Delta\vect u) 
	&= -\bar\Phi(\vect y) 
	+ \Delta\vect u^\top R\,\Delta\vect u \notag\\
	&\quad + (\vect u - \vect u^\star_{k+1} - \vect u_\theta)^\top Q_u\,
	(\vect u - \vect u^\star_{k+1} - \vect u_\theta), \label{eq_param_ell}\\
	V_{f,\theta}(\vect x_N) 
	&= \theta_V + \vect w_T^\top \boldsymbol{\sigma}_N, \label{eq_param_Vf}\\
	h_\theta(\vect x_k,\vect u_k) 
	&= \begin{bmatrix} 
		\underline{\vect x} - \vect s_\ell - \vect x_k \\ 
		\vect x_k - \overline{\vect x} - \vect s_u
	\end{bmatrix} \le \vect 0, \label{eq_param_h}
\end{align}
\end{subequations}
where $-\bar\Phi(\vect y)$ is the negated RTO economic profit 
(\eqref{eq_rto_param}), so that minimization of $\ell_\theta$ 
is equivalent to maximization of $\bar\Phi$, and 
$\boldsymbol{\sigma}_N$ is the terminal-slack decision variable with 
fixed weight $\vect w_T$. The RTO steady-state target 
$\vect u^\star_{k+1}$ enters the stage cost \eqref{eq_param_ell} 
as an \emph{input reference}, penalizing deviation of $\vect u_k$ from 
$\vect u^\star_{k+1}$ across the horizon; through the parameterized 
dynamics this indirectly pulls the predicted state trajectory toward 
the steady state associated with $\vect u^\star_{k+1}$. The terminal 
cost \eqref{eq_param_Vf} is deliberately kept to a scalar 
bias plus a slack penalty, playing the dissipativity role required 
by the terminal-cost ENMPC framework \citep{amrit2011economic} rather 
than an explicit terminal-state tracking role. The 
parameter partition of Algorithm~\ref{alg_main} therefore takes the 
concrete form
\begin{align}\label{eq_partition_instantiation}
	\begin{aligned}
		\vect\theta_V &= \theta_V \in \mathbb R, \\
		\vect\theta_m &\in \mathbb R^6\ (\text{as above}), \\
		\vect\theta_{\mathrm{RTO}} &= 
		\bigl(\vect\lambda_\theta^{\phi},\,\vect\lambda_\theta^{g},\,
		\vect\epsilon_\theta^{g}\bigr) \in \mathbb R^{21},
	\end{aligned}
\end{align}
together with $\vect\theta_c \in \mathbb R^{33}$, which stacks the 
entries of $R \in \mathbb{S}^3$, the reference offset 
$\vect u_\theta \in \mathbb R^3$, the entries of $Q_u \in \mathbb{S}^3$, 
and the state-bound offsets $\vect s = (\vect s_\ell,\vect s_u) \in \mathbb R^{12}$.
The offsets $\vect s$ shift the effective 
box in $h_\theta$: positive entries widen the box, negative entries 
narrow it, and the RL update is free to move them in either direction. 
The weight matrices $R,Q_u \in \mathbb{S}^3$ are projected onto the 
positive-semidefinite cone at each episode (via a small SDP) so that 
the parameterized ENMPC remains a convex-in-cost NLP. The RTO 
modifier vector $\vect\theta_{\mathrm{RTO}}$ collects the three groups 
of coefficients introduced in \eqref{eq_rto_param}.

\subsection{Dataset} \label{subsec_dataset}

The training dataset was collected in simulation with a sampling time of 
$T_s = 10\,\si{\second}$, matching the sampling time of the experimental 
rig. Each episode (one complete closed-loop run) 
consists of $1200$ steps ($200\,\si{\minute}$ of simulated closed-loop 
data). To 
improve generalization, data were collected under varying operating 
conditions, including changes in gas injection rates and deliberate 
adjustments of the reservoir valve openings (CV$101$--CV$103$) to 
emulate disturbances.

\subsection{Problem Formulation} \label{subsec_problem_formulation}

The optimal operation of the experimental rig is formulated as an 
economic optimization problem. The objective is to maximize the 
``oil'' revenue while respecting gas availability and injection 
constraints. Following \citep{matias2021steady}, the economic objective 
function is defined as
\begin{align}
	J = 20\,Q_{l,1} + 10\,Q_{l,2} + 30\,Q_{l,3},
\end{align}
where $Q_{l,i}$ denotes the liquid flow rate of well $i$. The weighting 
factors reflect hypothetical differences in hydrocarbon value among the 
three wells and are introduced for illustrative purposes.

The manipulated inputs are the gas injection flow rates:
\begin{align}
	\vect{u} = 
	\begin{bmatrix}
		Q_{\mathrm{gl},1} & Q_{\mathrm{gl},2} & Q_{\mathrm{gl},3}
	\end{bmatrix}^{\top},
\end{align}
where $Q_{\mathrm{gl},i}$ denotes the gas injected into well $i$. The constraint 
set $g_{\theta}$ in \eqref{eq_rto_param} and $h_\theta$ in 
\eqref{eq_enmpc_rl} include:
\begin{itemize}
	\item Lower and upper bounds on gas injection: 
	$Q_{gl,\min} = 1\,\si{s\liter\per\minute}$,
	$Q_{gl,\max} = 5\,\si{s\liter\per\minute}$.
	\item Gas availability constraint: 
	$Q_{\mathrm{gl},1} + Q_{\mathrm{gl},2} + Q_{\mathrm{gl},3} \leq 7.5\,\si{s\liter\per\minute}$.
\end{itemize}

Unless otherwise stated, the initial inputs are 
$Q_{\mathrm{gl},1} = Q_{\mathrm{gl},2} = Q_{\mathrm{gl},3} = 2.5\,\si{s\liter\per\minute}$.

\noindent\textbf{Plant-evaluated stage cost.} 
The stage cost $L$ used in the RL update 
\eqref{eq_q_learning_baseline_cost} is instantiated with 
$\ell_\theta$ set to the negative of the instantaneous economic 
objective,
\begin{align}\label{eq_ell_theta_case}
	\ell_\theta(\vect{x},\vect{u}) 
	= -\bigl(20\,Q_{l,1} + 10\,Q_{l,2} + 30\,Q_{l,3}\bigr),
\end{align}
so that minimizing $L$ is equivalent to maximizing $J$ while 
penalizing violations of $h_\theta$. Here $Q_{l,i}$ is the measured 
liquid flow rate of well $i$, i.e.\ a component of the plant output 
$\vect{y}$. The penalty vector $\vect{w} \in \mathbb{R}^{n_h}_{\ge 0}$ 
in \eqref{eq_q_learning_baseline_cost} weights state-bound violations 
of the parameterized inequalities $h_\theta$; in our runs we set 
all entries of $\vect{w}$ equal to $0.01$, small enough not to dominate 
the economic objective on typical trajectories yet large enough that 
persistent constraint violations are steadily driven out of the learned 
parameters \citep{gros2019data}. This makes explicit that a single 
scalar signal, the negative economic profit plus exterior-penalty on 
$h_\theta$, drives all RL updates.

\subsection{Offline Learning} \label{subsec_offline_learning}

For comparison, two different policies are generated: 
(i) the standard RLMPC approach in \eqref{eq_vf_parameterization} 
as presented in \citet{gros2019data}, and 
(ii) the proposed RTO--RLMPC approach following 
Algorithm~\ref{alg_main}. The learning procedure uses a discount factor 
of $\gamma = 0.99$. To ensure sufficient exploration, the optimal action is perturbed as
\begin{align}
	\vect{a} = \mathrm{sat}(\vect{u}^{\star}_0 + \vect{e},\, \underline{\vect{u}},\, \overline{\vect{u}}), 
	\quad \vect{e} \sim \mathcal{N}(\vect{0},\sigma_k^2 I),
\end{align}
where $\mathrm{sat}(\cdot,\underline{\vect{u}},\overline{\vect{u}})$ 
denotes elementwise saturation between the input bounds. The exploration standard deviation is decayed 
exponentially across episodes as
\begin{align}
	\sigma_k = \max\!\left(\sigma_{\min},\; \sigma_0 \cdot \rho^{k-1}\right),
\end{align}
with $\sigma_0 = 0.1$, $\sigma_{\min} = 0.01$, and $\rho = 0.995$, 
encouraging broad exploration early in training and narrowing the 
policy perturbations as the parameters converge. If a sampled action 
violates the total gas availability constraint $\sum_i u_i \leq 7.5$, 
rejection sampling is applied (up to $N_{\max}=50$ attempts) and the 
action is then rescaled onto the total-gas hyperplane, as in 
Algorithm~\ref{alg_main}.

The model parameters are projected after each update onto the box 
$[\underline{\vect{\theta}}_m,\overline{\vect{\theta}}_m]$ with 
$\underline{\vect{\theta}}_m = 0.5\,\vect{\theta}_{m,0}$ and 
$\overline{\vect{\theta}}_m = 1.5\,\vect{\theta}_{m,0}$ (i.e., 
$\pm 50\%$ around the initial values), which keeps the parameters in a 
physically meaningful range. The cost-weight matrices $R,Q$ are projected onto the positive semidefinite cone after each update by solving a small semidefinite program: $\min_{X\succeq 0} \|X - \tfrac{1}{2}(M+M^\top)\|_F^2$, with $M \in \{R,Q\}$. This returns the nearest PSD matrix to the (symmetrized) update and guarantees a well-posed ENMPC cost at every iteration.

Learning is terminated when the temporal-difference error $\tau$ in 
\eqref{eq_q_learning_td_update} converges to zero on average, or when the 
maximum number of episodes is reached. Since the parameters $\vect{\theta}$ 
are only locally valid, they may occasionally produce aggressive behavior. 
In practice, this can be mitigated by using parameter-dependent step sizes 
or applying simple first-order filters. The nominal learning rate was set 
to $\alpha = 10^{-2}$ for the terminal-cost parameter $\vect{\theta}_V$ 
and the input-move-penalty weight $R$ (part of $\vect{\theta}_c$), and 
$\alpha_m = 10^{-6}$ for the model parameters $\vect{\theta}_m$ and the 
remaining stage-cost/constraint parameters in $\vect{\theta}_c$ (the 
input-reference offset and the input-weighting matrix $Q_u$). The 
RTO-modifier parameters $\vect{\theta}_{\mathrm{RTO}}$ were updated once 
per episode with rate $\alpha_R = 10^{-2}$. If the NLP solver returns an 
infeasible status at any step, the previous feasible solution is reused 
and the parameter update is skipped for that step.

The resulting offline-trained policies are subsequently applied on the 
experimental rig for online validation (Section~\ref{sec_results}).

All optimization problems are formulated in CasADi 
\citep{Andersson2019} and solved with IPOPT 
\citep{wachter2006implementation}, with NLP sensitivities obtained via 
CasADi's automatic differentiation; the dynamic model is discretized 
with a fourth-order Runge--Kutta integrator over a prediction horizon 
of $N=10$ steps ($100\,\si{\second}$). Training runs for $200$ 
episodes of $1200$ steps each; on a standard desktop computer (Intel 
i7, $16\,\si{\giga\byte}$ RAM), each ENMPC and RTO solve takes 
approximately $0.2$--$0.5\,\si{\second}$, well within the 
$10\,\si{\second}$ sampling interval, and the full offline training 
completes in roughly one day of wall-clock time, including sensitivity 
computations and projection steps.
\section{Results} \label{sec_results}

Before implementing the learned policies on the experimental rig, the 
controller was first validated on a high-fidelity MATLAB plant 
simulator (the controller itself is implemented in CasADi, as described 
in Section~\ref{sec_case_study}; code available online, see the Data 
Availability statement). The simulation studies predicted profit 
improvements of the same order as those subsequently observed on the 
rig, and the experimental gains reported below lie within the 
variability of the simulated runs.

We compare three control strategies: (i) a nominal ENMPC, (ii) an RLMPC 
trained according to \citep{gros2019data}, and (iii) the proposed 
RTO--RLMPC framework (Algorithm~\ref{alg_main}). A standard 
modifier-adaptation (MA) baseline is not included because MA requires 
explicit plant gradient estimates, which demand costly perturbation 
experiments on the rig; the proposed method eliminates this requirement 
entirely. The interested reader is referred to \citet{turan2023experimental} 
for MA results on the same experimental setup.

\begin{figure}[]
	\centering
	\includegraphics[width=\columnwidth]{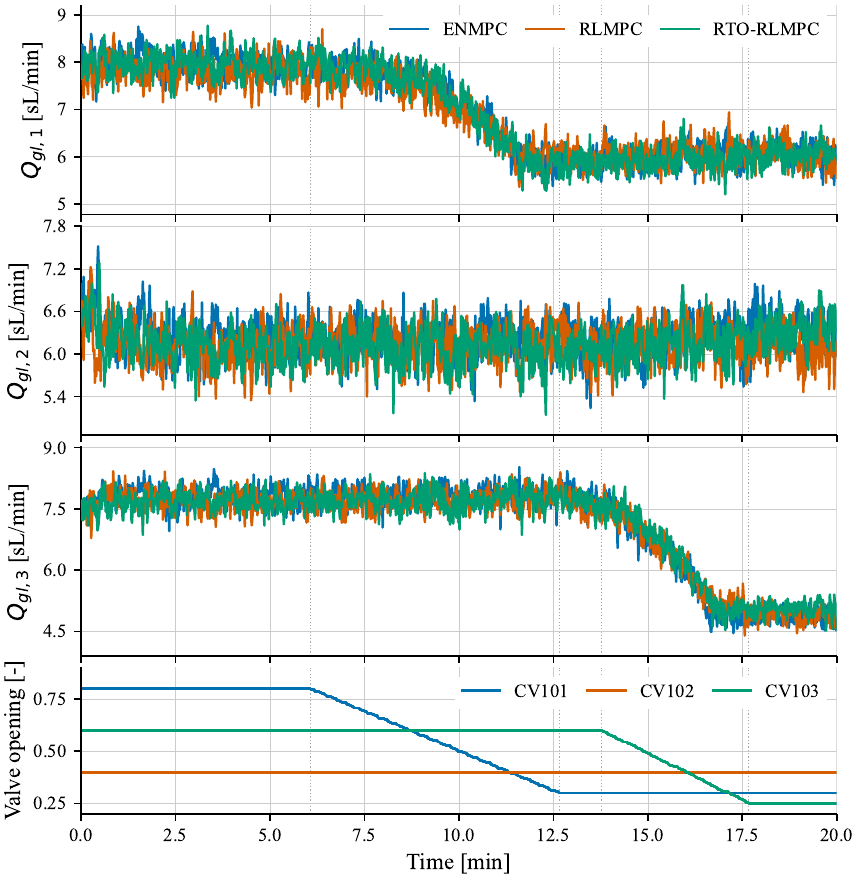}
	\caption{Gas-lift injection rates $Q_{\mathrm{gl},i}$ for the three wells under ENMPC, 
		RLMPC, and the proposed RTO--RLMPC. The top three panes show the 
		manipulated inputs, while the bottom pane depicts the disturbance profile 
		(CV$101$--CV$103$ openings) emulating declining production in wells~$1$ and~$3$.}
	\label{fig_gas_input}
\end{figure}

\begin{figure}[]
	\centering
	\includegraphics[width=\columnwidth]{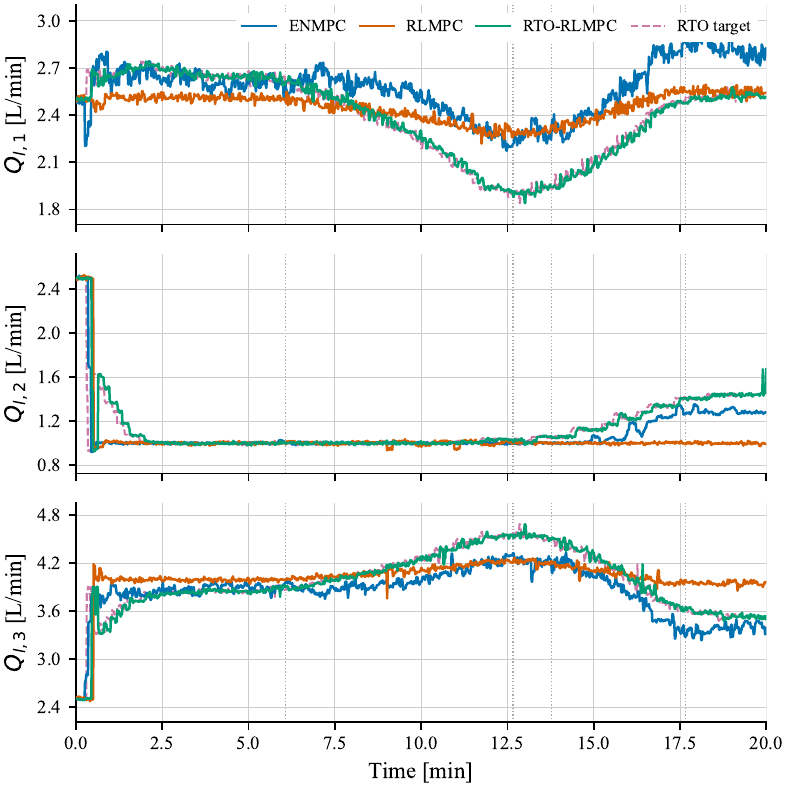}
	\caption{Liquid flow rates $Q_{l,i}$ for the three wells under ENMPC, RLMPC, 
		and the proposed RTO--RLMPC. Dashed lines indicate the optimal 
		steady-state flow rates computed by the RTO layer. Well~$3$ carries the 
		highest economic weight and shows the most pronounced improvement under 
		RTO--RLMPC.}
	\label{fig_liquid_flowrate}
\end{figure}

\begin{figure}[]
	\centering
	\includegraphics[width=\columnwidth]{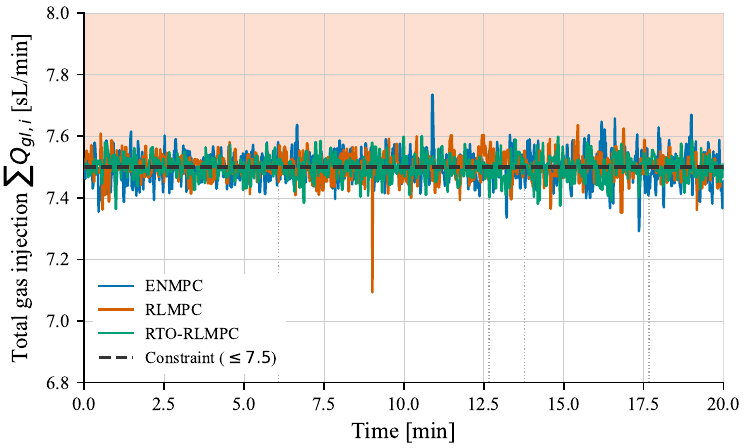}
	\caption{Constraint satisfaction for the total gas injection 
		$Q_{\mathrm{gl},1} + Q_{\mathrm{gl},2} + Q_{\mathrm{gl},3} \leq 7.5\,\si{s\liter\per\minute}$ under 
		ENMPC, RLMPC, and the proposed RTO--RLMPC, based on experimental results.}
	\label{fig_constraints}
\end{figure}

\subsection{Experimental Results} \label{subsec_exp_results}
\begin{figure}[]
	\centering
	\includegraphics[width=\columnwidth]{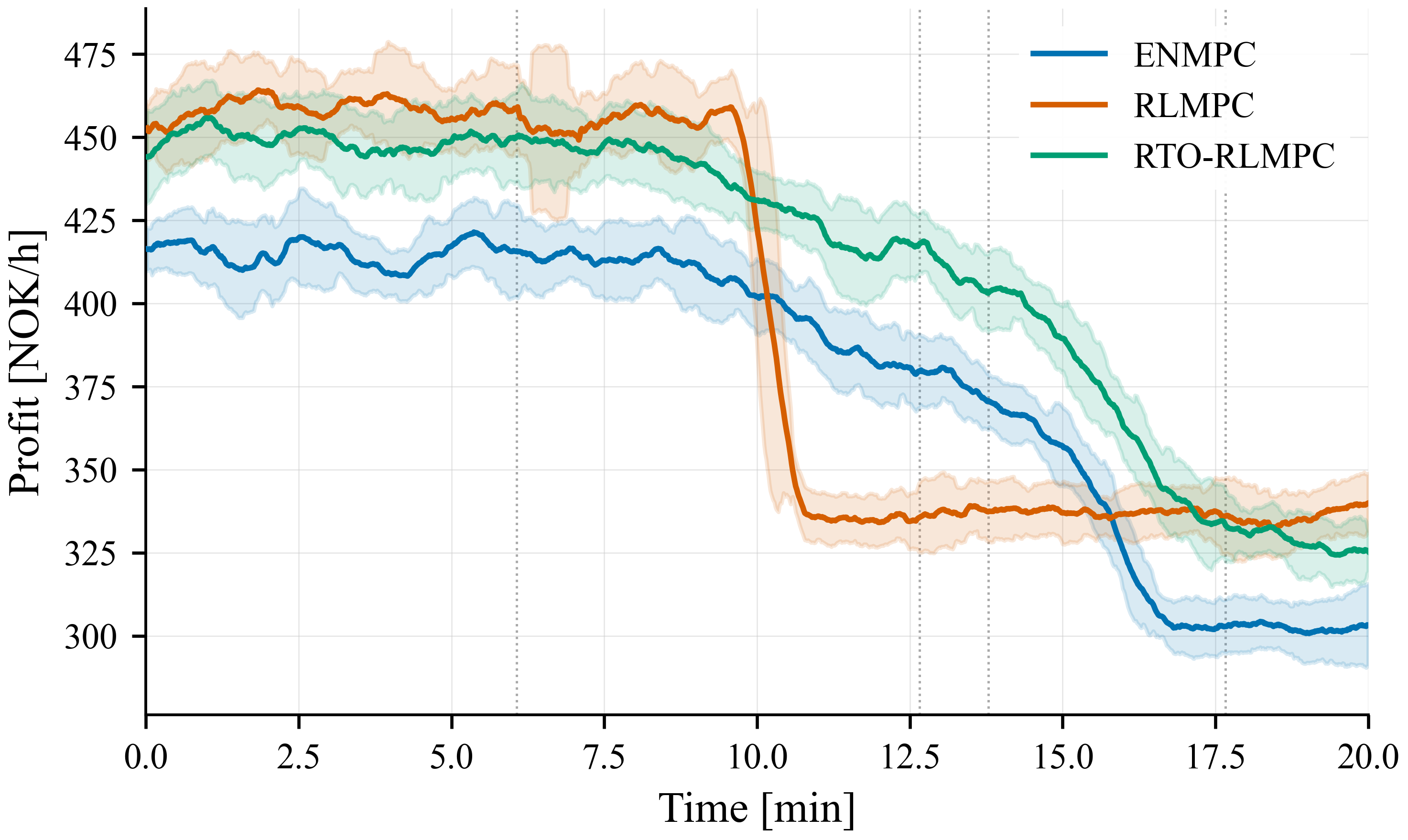}
	\caption{Profit comparison under ENMPC, RLMPC, and the proposed 
		RTO--RLMPC. Signals are smoothed with a $60\,\si{\second}$ moving 
		average; shaded bands represent local variability across three 
		repeated experiments.}
	\label{fig_profit}
\end{figure}

The three controllers were tested under a disturbance scenario designed 
to emulate declining well production. Each experiment lasted 
$20\,\si{\min}$, with disturbances introduced by varying the openings of 
CV$101$--CV$103$ while keeping the pump outlet pressure $P_{\mathrm{pump}}$ 
constant. The disturbance profile is shown in the bottom pane of 
Fig.~\ref{fig_gas_input}.

The top three panes of Fig.~\ref{fig_gas_input} show the manipulated 
variables $Q_{\mathrm{gl},i}$. During the first stage ($t=0$--$6\,\si{\min}$), 
well~3 receives the largest gas injection due to its higher weight in the 
economic objective, while well~2 remains at the minimum injection level 
$Q_{gl,\min}=1\,\si{s\liter\per\minute}$. In the second stage 
($t=6$--$14\,\si{\min}$), as CV$101$ gradually closes, the allocation 
shifts further toward well~3. Toward the end of the run, gas injection in 
well~2 begins to increase as CV$101$ and CV$103$ become more restrictive, 
yet the order $Q_{\mathrm{gl},3}>Q_{\mathrm{gl},1}>Q_{\mathrm{gl},2}$ is maintained.

Overall, Fig.~\ref{fig_gas_input} confirms that all three controllers 
(nominal ENMPC, RLMPC, and RTO--RLMPC) produce consistent qualitative 
behavior. The gas-injection signals are affected by a high noise-to-signal 
ratio, which makes it difficult to firmly establish input-level trends 
from Fig.~\ref{fig_gas_input} alone; with this caveat, RTO--RLMPC appears 
to respond more rapidly to disturbances and produce larger input 
reallocations, while RLMPC appears to adapt more slowly, potentially 
reflecting the absence of a dedicated RTO layer for steady-state 
correction. This trend is corroborated more clearly by the resulting 
liquid flow rates and profit, discussed next.

Fig.~\ref{fig_liquid_flowrate} shows the resulting liquid flow rates 
$Q_{l,i}$ for the three wells. Since the profit is a weighted sum of 
these flow rates, this figure directly illustrates the economic 
performance of each controller. The dashed lines indicate the optimal 
steady-state flow rates computed by the RTO layer. The proposed 
RTO--RLMPC achieves liquid flow rates that track the RTO targets more 
closely than the nominal ENMPC, particularly during the second and third 
disturbance phases when valve openings change. Well~3, which carries the 
highest economic weight, shows the most pronounced improvement.

Constraint enforcement is illustrated in Fig.~\ref{fig_constraints}. 
By construction of the projection step in Algorithm~\ref{alg_main}, the 
commanded total gas injection satisfies 
$Q_{\mathrm{gl},1}+Q_{\mathrm{gl},2}+Q_{\mathrm{gl},3}\le 
7.5\,\si{s\liter\per\minute}$ at every step; the measured total shown in 
the figure occasionally lies slightly above the bound during 
disturbance transitions, with magnitude comparable to the flowmeter 
noise floor. Over the $1200$ samples of each run, the mean absolute 
deviation of the measured total gas injection from the 
$7.5\,\si{s\liter\per\minute}$ limit is 
$0.037$, $0.031$, and $0.028\,\si{s\liter\per\minute}$ for ENMPC, RLMPC, 
and RTO--RLMPC, respectively, and the largest observed exceedance of the 
limit is $0.235$, $0.136$, and $0.100\,\si{s\liter\per\minute}$. 
The proposed RTO--RLMPC thus operates closest to the constraint on 
average and exhibits the smallest worst-case exceedance among the three 
controllers, extracting more economic benefit while remaining the 
best-behaved with respect to the shared-input limit.

Finally, Fig.~\ref{fig_profit} compares the resulting profit. The signals 
are smoothed using a $60\,\si{\second}$ moving average to reduce the 
effect of noise. Both RL-based controllers achieve higher economic 
performance than the nominal ENMPC: RLMPC improves profit by 
approximately $4.7\%$, and the proposed RTO--RLMPC yields an $8.6\%$ 
increase relative to ENMPC. The experiments were repeated three times 
for each controller; the shaded bands in Fig.~\ref{fig_profit} represent 
the local standard deviation across runs, confirming that the observed 
improvements are reproducible.

\begin{figure}[]
	\centering
	\includegraphics[width=\columnwidth]{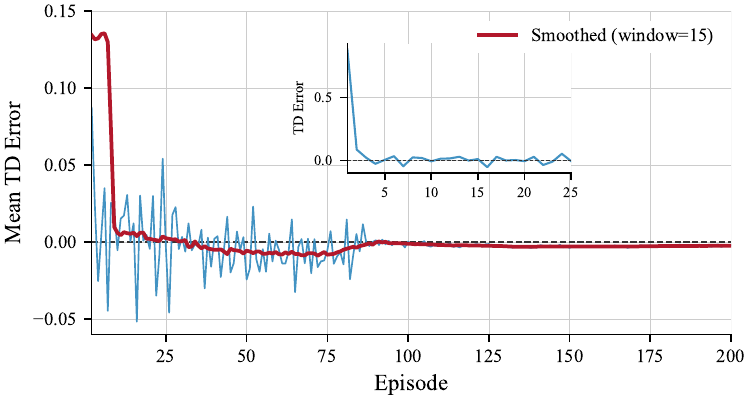}
	\caption{Episode-averaged temporal-difference error $\bar{\tau}$ 
		over $200$ episodes of offline training. Convergence toward zero 
		indicates consistency between the learned Q-function and the 
		observed cost-to-go.}
	\label{fig_td_convergence}
\end{figure}

\begin{figure}[]
	\centering
	\includegraphics[width=\columnwidth]{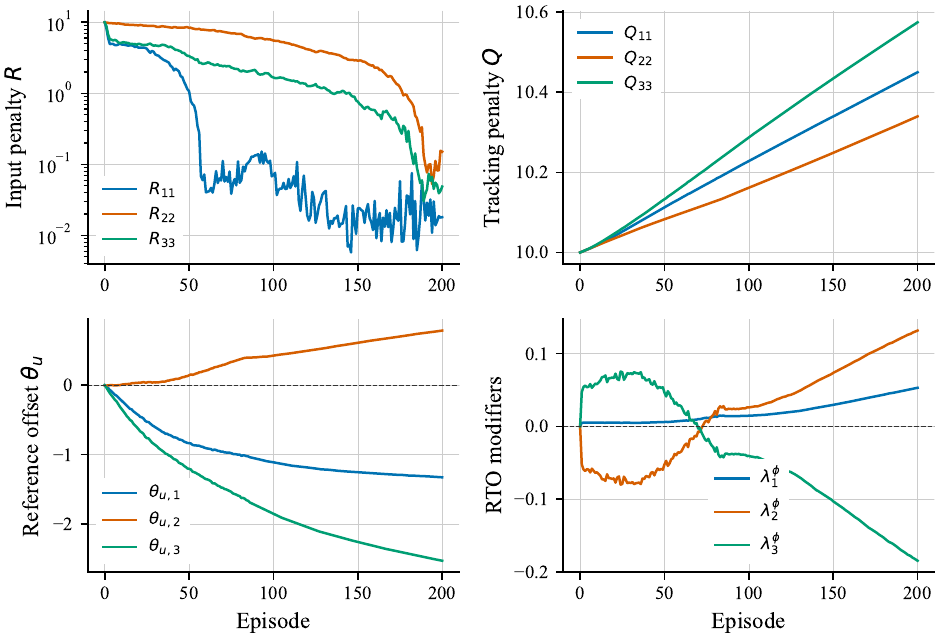}
	\caption{Evolution of cost weights ($R$, $Q$ diagonals), reference 
		offsets $\theta_u$, and RTO modifier terms over training episodes. 
		The movement penalty $R$ decreases by two orders of magnitude, 
		enabling more responsive control.}
	\label{fig_cost_evolution}
\end{figure}

\begin{table}[]
	\centering
	\caption{Comparison of nominal (initial), learned, and reference 
		model parameters after $200$ episodes of offline training. 
		Reference values were identified independently via a separate 
		calibration experiment and represent the best available estimate 
		of the true dynamics, not exact ground truth.}
	\label{tab_params}
	\begin{tabular}{lccc}
		\hline
		Parameter & Initial & Learned & Reference \\
		\hline
		$\theta_{\mathrm{res},1}$ & $0.315$ & $0.342$ & $0.415$ \\
		$\theta_{\mathrm{res},2}$ & $0.063$ & $0.135$ & $0.163$ \\
		$\theta_{\mathrm{res},3}$ & $0.164$ & $0.226$ & $0.264$ \\
		$\theta_{\mathrm{top},1}$ & $0.530$ & $0.536$ & $0.630$ \\
		$\theta_{\mathrm{top},2}$ & $0.571$ & $0.572$ & $0.671$ \\
		$\theta_{\mathrm{top},3}$ & $0.622$ & $0.619$ & $0.722$ \\
		\hline
		$R_{11}$  & $10.0$ & $0.018$ & -- \\
		$R_{22}$  & $10.0$ & $0.151$ & -- \\
		$R_{33}$  & $10.0$ & $0.049$ & -- \\
		\hline
	\end{tabular}
\end{table}

\subsection{Learning Convergence} \label{subsec_learning_convergence}
Fig.~\ref{fig_td_convergence} shows the episode-averaged 
temporal-difference error $\bar{\tau}_k = \tfrac{1}{T}\sum_{i=1}^T \tau_i$ 
as a function of the episode index. The TD error converges to zero within 
approximately $150$ episodes, indicating that the Q-function and value 
function become consistent with the observed stage costs and transitions.

The model parameters $\vect{\theta}_m$ move toward their reference 
plant values (Table~\ref{tab_params}) along the directions that most 
affect the closed-loop economic objective. This selective convergence 
follows directly from the structure of the update rule 
\eqref{eq_q_learning_theta_update}: each component of $\vect{\theta}_m$ 
is updated in proportion to the TD error times the corresponding 
partial derivative $\partial Q_\theta/\partial \theta_m$ 
\eqref{eq_enmpc_sensitivities}, so a parameter direction along which 
$Q_\theta$ is weakly sensitive to $\theta_m$ receives a correspondingly 
weak learning signal, regardless of how much closed-loop data are 
collected. This mirrors the persistence-of-excitation requirement in 
classical system identification: only those directions that are 
sufficiently ``excited'' economically, i.e., that measurably change the 
action--value function, are well identified from the available 
trajectories; directions weakly coupled to the cost remain only 
partially identified even after the TD error has converged. The cost 
weights and RTO modifier evolution are shown 
in Fig.~\ref{fig_cost_evolution}; notably, the movement penalty $R$ 
decreases significantly, allowing the controller to respond more 
aggressively to disturbances.

We further observe a correlation between the convergence of 
$\vect{\theta}_m$ and the evolution of the RTO modifier 
$\vect{\lambda}_\theta^\phi$ entering the RTO objective 
\eqref{eq_rto_param}: once $\vect{\theta}_m$ stabilizes, 
$\vect{\lambda}_\theta^\phi$ keeps 
growing in magnitude (Fig.~\ref{fig_cost_evolution}) instead of 
settling. This is consistent with the modifiers absorbing whatever 
residual plant--model mismatch the now-frozen model parameters can no 
longer explain: if the active total-gas constraint in \eqref{eq_rto_param} 
leaves a direction of $\vect{\lambda}_\theta^\phi$ along which the 
sensitivity $\partial \vect{u}^\star/\partial\vect{\theta}_{\mathrm{RTO}}$ 
vanishes, the TD-error-driven update \eqref{eq_q_learning_theta_update} 
integrates unopposed and $\vect{\lambda}_\theta^\phi$ drifts without 
bound, even though $\vect{u}^\star$ itself remains well-behaved. This 
mirrors the known non-uniqueness of modifiers under active constraints 
in the modifier-adaptation literature, and is consistent with the 
persistent lack of input convergence reported by 
\citet{matias2021online, turan2023experimental} on the same rig, where 
a similarly flat direction in the RTO sensitivity map left the modifier 
update unopposed. Unlike $\vect{\theta}_m$, which is kept within 
a physically motivated box $[\underline{\vect{\theta}}_m,\overline{\vect{\theta}}_m]$ 
(Section~\ref{subsec_offline_learning}), $\vect{\theta}_{\mathrm{RTO}}$ 
is currently unconstrained; a regularization or projection step 
analogous to the model-parameter box constraint would likely prevent 
this unbounded growth, and we leave a principled design of such a 
safeguard to future work.

Table~\ref{tab_params} summarizes the initial, learned, and reference 
parameter values. The reservoir valve coefficients $\theta_{\mathrm{res},i}$ 
show the largest relative change, consistent with the fact that the 
main disturbances originate from the reservoir valves.
The top-valve parameters $\theta_{\mathrm{top},i}$ show smaller relative changes, 
suggesting that the economic cost is less sensitive to these parameters 
and that they are harder to identify from closed-loop data alone. 
The partial convergence of $\vect{\theta}_m$ to the reference values in 
Table~\ref{tab_params} is therefore consistent with the converged TD 
error in Fig.~\ref{fig_td_convergence}: training was not stopped 
prematurely, but rather the closed-loop data are not sufficiently 
informative to identify the parameters along directions that have a 
weak effect on the closed-loop economic objective.
A natural question is whether this model adaptation is essential to the 
framework's performance, or whether cost-parameter tuning alone could 
achieve similar results.

\subsection{Ablation: Role of Model Learning} \label{subsec_ablation}
To assess the contribution of model learning, we run an ablation in 
which the model parameters are frozen at their initial (misspecified) 
values ($\alpha_m = 0$), while RTO and cost-parameter updates proceed 
as usual, and compare the economic objective per episode for the two 
configurations.

Both start from identical initial conditions and achieve similar 
performance in the early episodes. As training progresses, the 
proposed method maintains a steady economic objective near 
$447 \times 10^{3}$, whereas the variant without model learning 
degrades to $436 \times 10^{3}$, a $2.5\%$ reduction. This decline in 
the ablation curve reflects the agent settling on more conservative 
cost weights when model learning is disabled: starting from the same 
initial $R_0=\mathrm{diag}(10,10,10)$, the $R$-matrix norm converges to 
$\approx 0.19$ for the proposed method (consistent with the learned 
values in Table~\ref{tab_params}) but to $2.44$ for the ablation, 
indicating that without model updates the agent compensates by 
remaining far more conservative, trading a small amount of economic 
performance for improved constraint satisfaction. 

More importantly, the mean TD error fails to converge without model 
learning ($|\bar{\tau}| = 0.015$ vs.\ $0.001$ for the proposed method), 
indicating that the cost parameters alone cannot compensate for 
persistent model mismatch. This suggests that adapting the internal 
model is important for the MPC to serve as a reliable function 
approximator for the $Q$-function.

\section{Conclusion} \label{sec_conclusion}

This work proposed a reinforcement-learning-based framework that 
integrates real-time optimization (RTO) and economic nonlinear MPC 
(ENMPC) through modifier-adaptation concepts. A parameterized ENMPC is 
augmented with modifier terms from an RTO layer that drives the system 
to the true steady-state optimum, which is embedded in the ENMPC 
stage cost as a reference. This removes the need for explicit steady-state gradient estimation typically required in modifier adaptation.

The framework was experimentally validated on a laboratory-scale 
gas-lifted oil-well network. The proposed RTO--RLMPC achieved an 
average profit improvement of approximately $8.6\%$ over nominal ENMPC 
and tracked the economically optimal steady states more closely, 
particularly for the highest-value well. The temporal-difference 
error converged within roughly $150$ episodes; the learned model 
parameters moved toward reference plant values, independently 
identified from experimental data, along the directions that most 
influence the economic objective, while directions weakly coupled to 
the closed-loop cost remained only partially identified 
\citep{gros2019data}.

Compared to standard modifier adaptation, RTO--RLMPC does not require explicit plant gradient 
estimation, which is often the primary bottleneck in industrial 
deployment. Compared to model-free RL with neural-network policies, 
the MPC-based parameterization retains an interpretable, 
constraint-aware structure: for any \emph{fixed} parameter vector 
$\vect{\theta}$, the standard dissipativity-based ENMPC stability 
argument \citep{amrit2011economic} applies under the usual assumptions 
on the terminal cost, and the learned quantities (cost weights, 
modifier terms, and reference offsets) carry a direct physical 
interpretation. Formal 
transfer of these certificates to the \emph{learning phase}, in which 
model, cost, constraints and terminal cost evolve online, 
remains an open theoretical question; in this work it is addressed 
pragmatically through a two-timescale update strategy (fast ENMPC 
updates per step, slow RTO updates per episode) and empirically 
verified on the rig.

\textbf{Limitations and future work.} 
Performance is fundamentally limited by the richness of the chosen 
parameterization $(f_\theta,\ell_\theta,V_{f,\theta},h_\theta)$: 
insufficient flexibility here caps the achievable performance regardless 
of the amount or duration of training. The offline training data 
(Section~\ref{subsec_dataset}) also cover a limited set of operating 
conditions and disturbance patterns, so generalization to unseen 
scenarios remains open, and the cubic scaling of the NLP sensitivity 
computation with the number of decision variables will require 
decomposition or approximate sensitivities for larger networks. Finally, 
the unbounded growth of $\vect{\theta}_{\mathrm{RTO}}$ noted in 
Section~\ref{subsec_learning_convergence} indicates that a 
regularization or projection step, analogous to the model-parameter box 
constraint, is needed and is left for future work.
\appendix
\renewcommand{\theequation}{A.\arabic{equation}}
\setcounter{equation}{0}
\section{Model Equations}
\label{appendix_model_eq}

The dynamic model, adapted from \citet{marchetti2014steady}, comprises 
mass balances and pressure/flow relations for three identical wells. Mass balances:
\begin{subequations}\label{eq_mass_balance}
	\begin{align}
		\dot{m}_{g} &= w_g - w_{g,out}, \quad
		\dot{m}_{l} = w_l - w_{l,out},
	\end{align}
\end{subequations}
where $m_{g,l}$ are holdups, $w_{g,l}$ injection and inflow, and $w_{g,l,out}$ outlet flows. 
Reservoir outflow:
\begin{equation}\label{eq_liq_flow}
	w_l = v_o \theta_{\mathrm{res}}\sqrt{\rho_l\,(P_{\mathrm{pump}} - P_{\mathrm{bi}})}.
\end{equation}
Pressure before injection using Darcy--Weisbach:
\begin{equation}\label{eq_pressDrop}
	P_{\mathrm{bi}} = P_{\mathrm{rh}} + \rho_{\mathrm{mix}} g \Delta h 
	+ \frac{128 \mu_{\mathrm{mix}}(w_g + w_l)L}{\pi \rho_{\mathrm{mix}} D^4}.
\end{equation}
Mixture density: $\rho_{\mathrm{mix}} = (m_g + m_l)/V_{\mathrm{total}}$. 
Total outlet flow: 
\begin{equation}\label{eq_flow}
	w_{\mathrm{total}} = \theta_{\mathrm{top}}\sqrt{\rho_{\mathrm{mix}}\,(P_{\mathrm{rh}} - P_{\mathrm{atm}})}.
\end{equation}
\section*{Declaration of Competing Interests}
The authors declare that they have no known competing financial interests or 
personal relationships that could have appeared to influence the work reported 
in this paper.

\section*{Funding}
This work was supported by the Research Council of Norway under the IKTPLUSS 
program (Project number $299585$) and by SINTEF internal research funds.

\section*{Data Availability}
The simulator code, experimental rig model parameters, and scripts used to 
reproduce the results of this study are openly available at 
\url{https://github.com/saketadhau/rl-rto-enmpc}.

\section*{Declaration of Generative AI and AI-Assisted Technologies 
in the Manuscript Preparation Process}
During the preparation of this work the authors used GitHub Copilot 
in order to assist with consistency checks, 
and language polishing. After using this tool, the authors reviewed 
and edited the content as needed and take full responsibility for 
the content of the published article.
\printcredits
\bibliographystyle{elsarticle-num-names}
\bibliography{mainRigbib}
\end{document}